\documentstyle[11pt,newpasp,twoside,epsf]{article}
\input epsf
\def\edcomment#1{\iffalse\marginpar{\raggedright\sl#1\/}\else\relax\fi}
\reversemarginpar

\begin{document}
\title{Non-LTE Effects in Berylium Abundances}
 \author{T. P. Idiart}
\affil{Universidade de S\~ao Paulo, IAG, Depto. de Astronomia \\
Av. Miguel Stefano 4200, S\~ao Paulo 01065-970, Brazil}
\author{F. Th\'evenin}
\affil{Observatoire de la C\^{o}te d'Azur \\ 
B.P. 4229, 06304 Nice Cedex 4, France}

\begin{abstract}
In this work we analyze the beryllium-iron chemical diagram from the 
point of view of non-LTE effects. Be abundances were re-calculated by 
considering non-LTE corrections in ionization equilibrium (logg) and Fe 
abundances ([Fe/H]). These corrections seem do not affect the linear 
relation between Be-Fe for metal-poor stars already found in the literature 
for LTE derived abundances.
\end{abstract}

\section{Introduction}
The analysis of the trends of abundances of light elements with 
respect to [Fe/H] for the oldest metal-poor stars is a direct way to 
provide some clues on their production mechanism and evolution.

In a recent work on non-LTE effects in iron abundances, Th\'evenin \& Idiart 
(1999) (TI99) obtained that for metal-poor dwarf stars Fe abundances 
([Fe/H]) are affected by significant non-LTE effects and, moreover, surface 
gravities (logg) derived by LTE analysis also need corrections. 
This logg corrections should be crucial for beryllium abundances determination, 
since Be II resonance lines normally used to estimate Be abundances are much 
sensitive to this stellar parameter. 

In this work we examine the consequences of non-LTE corrections to 
logg and [Fe/H] for logN(Be/H) vs. [Fe/H] (or Be-Fe) diagram. In section 
2 we present a short summary of our results obtained in 
TI99 for Fe and in section 3 the results for Be abundances .

\section{Non-LTE Corrections for [Fe/H] and logg}

TI99 performed statistical equilibrium calculations for Fe I and 
Fe II to estimate non-LTE effects in iron abundances. 
The main results are showed in figure 1 (see TI99 for details).

\begin{figure}[h]
\plotfiddle{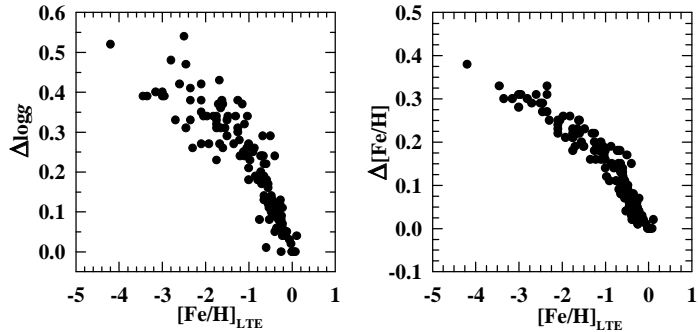}{2.0 cm}{0}{50}{50}{-170}{-305}
\vspace{1.8 cm}
\caption{Amplitude of non-LTE logg and [Fe/H] corrections in function  
of LTE [Fe/H]  for  136 subgiant to subdwarf stars.}
\end{figure}

\section{Results for Be/H}

We re-estimate N(Be/H) abundances for 21 stars also analised by Boesgaard et 
al.(1999) using $\rm T_{eff}$, logg and [Fe/H] given by TI99. Be abundances were 
calculated assuming LTE conditions, since the non-LTE corrections are negligible 
for Be II lines considered here 
($\rm \lambda \lambda$ 3130 and 3131), as demonstrated by Garcia Lopez et al. 
(1995), for example. Figure 2 shows our results.

\begin{figure}[h]
\plotfiddle{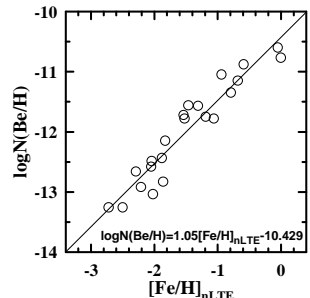}{1.0 cm}{0}{40}{40}{-170}{-185}
\vspace{2.8 cm}
\caption{Derived Be abundances vs. non-LTE corrected [Fe/H].} 
\end{figure}

We conclude that for the range of metal-poor objects  -3 
$\rm <$ [Fe/H] $\rm <$ -1.5, 
non-LTE corrections for [Fe/H] compensate changes in 
Be abundances (as result of logg corrections) in the Be-Fe diagram, 
recovering the same linear behavior of LTE derived abundances (Boesgaard et 
al.1999). Similar results are found for Boron (see Primas 1999, this colloquium).

\end{document}